\documentclass[12pt]{article}
\usepackage{amssymb}
\usepackage{amsmath}
\usepackage{graphicx}
\usepackage{bm}
\newcommand{\mathsym}[1]{{}}

\usepackage{graphicx}
\usepackage{rotating}
\usepackage{color}
\usepackage{cancel}

\newcommand{\bra}{\begin{array}}
\newcommand{\era}{\end{array}}
\newcommand{\beq}{\begin{equation}}
\newcommand{\eeq}{\end{equation}}
\newcommand{\beqar}{\begin{eqnarray}}
\newcommand{\eeqar}{\end{eqnarray}}

\newcommand{\be}{\begin{equation}}
\newcommand{\ee}{\end{equation}}
\newcommand{\bea}{\begin{eqnarray}}
\newcommand{\eea}{\end{eqnarray}}
\newcommand{\bd}{\begin{displaymath}}
\newcommand{\ed}{\end{displaymath}}

\newcommand{\h}{ \hbar}

\newcommand{\op }{\oplus_{\al}}
\newcommand{\om }{\ominus_{\al}}

\newcommand{\lb }{ \left (}
\newcommand{\rb }{ \right )}

\newcommand{\s }{\sigma}

\newcommand{\ta }{ \theta}

\newcommand{\tb }{ \Theta}

\newcommand{\al }{\alpha}

\begin{document}

\vspace{20pt}

\begin{center}

{\LARGE \bf Quantum mechanics on a circle with a finite number of $\al$-uniformly distributed points

\medskip
 }
\vspace{15pt}

{\large  Won Sang Chung${}^{1}$, \,Ilyas Haouam${}^{2,\dag}$
and Hassan Hassanabadi${}^{3}$
}

\vspace{15pt}
{\sl ${}^{1}$Department of Physics and Research Institute of Natural Science,\\
 College of Natural Science,\\
 Gyeongsang National University, Jinju 660-701, Korea}\\
{\sl ${}^{2}$ Laboratoire de Physique Mathematique et de Physique Subatomique (LPMPS), Universite Freres Mentouri, 25000, Constantine, Algeria}\\

{\sl ${}^{3}$ Faculty of physics, Shahrood University of Technology, Shahrood, Iran }

\vspace{5pt}
E-mail:
{${}^{\dag}$ ilyashaouam@live.fr (Corresponding author.)}

\vspace{10pt}

\end{center}

\begin{abstract}

In this paper{\color{black},} quantum mechanics on a circle with {\color{black}finite} number of $\al$-uniformly distributed points is discussed. The angle operator and translation operator {\color{black}are} defined. {\color{black}Using}  discrete angle representation, two types of discrete angular momentum operators and Hermitian Hamiltonian on a circle with $d$ $\al$-distributed discrete angles are constructed. The energy levels are computed for a free particle on a circle where the wave function is defined in the $d$ $\al$-distributed discrete angles.

\end{abstract}

 \today

\section{Introduction}

The quantum mechanics on a circle is characterized by the position operator measuring the angle and angular momentum operator playing a role of translation on a {\color{black}circle [1-8]}. The angle is real-valued and it is restricted in $[0, 2\pi)$ from the periodic properties.

Schwinger [9] presented a most interesting model where the position space
and the momentum space  consist of finite number of points and are periodic. This was referred to as the quantum mechanics in a finite Hilbert space [9-12].

The idea of finite Hilbert space was adopted in constructing the quantum mechanics on a circle [1-8] with a finite number of points, where the discrete angle was introduced for describing the position of a particle on a circle and they are uniformly distributed on a circle. Thus, the angular momentum is given by the finite difference with respect to discrete angles.
The finite quantum mechanics on a circle becomes the continuous quantum mechanics on a circle from the continuous limit. Then, the finite difference with respect to discrete angles becomes the derivative  with respect to a continuous angle in this limit.

Bonatsos et al. [13] have considered the position and momentum operators for the $q$-deformed oscillator with $q$ being a root of unity. They have shown that the phase space of this oscillator has a lattice structure, which is a
non-uniformly distributed grid.

In this paper we consider another type of the non-uniformly distributed grid, where we consider the discrete angles are described in terms of the $\al$-addition [14]. Using these discrete angles, we construct the quantum mechanics on a circle with a finite number of $\al$-uniformly distributed points.

\section{$\al$-uniformly distributed angles  on a circle}

Now let us consider the quantum mechanics on a circle with finite points. We
 consider the case that   $d$ points
 \be
 \{ \ta_0, \ta_1, \cdots, \ta_{d-1}\}
 \ee
 on a circle are allowed and they are not uniformly distributed. Here we set $\ta_0=0$ and
 \be
 \ta_d \equiv \ta_0~~~ (mod ~ 2 \pi )
 \ee
 and
 \be
 \ta_0 < \ta_1 < \cdots <  \ta_{d-1} < 2 \pi
 \ee
 In this paper we consider the discrete angles are given in terms of the deformed addition called $\al$-addition,
 \be
  \ta_{n+1} \om \ta_n =\s, ~~~or~~~\ta_{n+1} = \ta_n \op \s, ~~~\s>0,
 \ee
 where $\al$-addition and {\color{black} $\al$-subtraction [6] are defined as}
  \bea
a\op b &=& |a |a|^{\al-1}+ b|b|^{\al-1}|^{1/\al-1} ( a |a|^{\al-1}+ b|b|^{\al-1}) \\
a\om b &=& |a |a|^{\al-1}- b|b|^{\al-1}|^{1/\al-1} ( a |a|^{\al-1}- b|b|^{\al-1}),
\eea
{\color{black}with 
$a\otimes_{\al} b =a\, b$ and 
$a\oslash_{\al} b = \frac{a}{b}$ ($\alpha$-product and $\alpha$-division )}. Solving Eq.(4), we get
 \be
 \ta_n =n^{\frac{1}{\al}}\s , ~~~~ n=0, 1, \cdots, d-1
 \ee
 If we demand that $d$-th position is the same as the $0$-th position we have
 \be
 \ta_d = d ^{\frac{1}{\al}}\s=2 \pi,
 \ee
 which gives
 \be
 \s  = \frac{2 \pi}{d ^{\frac{1}{\al}}}
 \ee
 Thus, the relation (4) is replaced with
 \be
 \ta_{n+1} \om \ta_n =\frac{2 \pi}{d ^{\frac{1}{\al}}},
 \ee
 and
 \be
 \ta_n = 2\pi \lb \frac{n}{d}\rb^{\frac{1}{\al}}
  \ee
 In the limit $d \rightarrow \infty$, the lattice $\{ \ta_n | n=0, 1, \cdots, d-1\}$ becomes the semi open interval $0 \le \ta < 2 \pi$.

 \section{Angle operator,  translation operator and discrete angle representation}

 Now let us discuss a finite quantum mechanics on a circle $d$ $\al$-distributed discrete angles.
 The Hermitian  angle operator $\tb$ acts on the angle eigenvectors denoted by $|n\rangle, n = 0, 1, 2,\cdots, d-1$ such
that
\be
\tb |n\rangle =\ta_n |n\rangle = 2\pi \lb \frac{n}{d}\rb^{\frac{1}{\al}} |n\rangle
\ee
We know that Eq.(12) is not well-defined because the operator $\tb$  is defined modulo $2\pi$. Thus, we introduce a new operator defined as
\be
V = e_{\al} ( i \xi \tb),
\ee
where $\xi>0$ is real and will be fixed later, and the $\al$-exponential [15] is defined as
\be
e_{\al}(z) = e^{ |z|^{\al-1} z},
\ee
{\color{black}with $-\infty < z < +\infty$. } The $\al$-exponential obeys [15]
\be
e_{\al} ( z \op w ) = e_{\al}(z) e_{\al}(w)
\ee
\be{\color{black}
e_{\al} ( z \om w ) = e_{\al}(z)\oslash_{\al} e_{\al}(w)}\nonumber
\ee
Acting the operator $V$ on $|n\rangle$, we get
\be
V |n\rangle = e_{\al} ( i \xi \ta_n ) |n\rangle =e^{ i \xi^{\al} \ta_n^{\al} } |n\rangle =
e^{ i \xi^{\al} ( 2 \pi)^{\al} \frac{n}{d}  }|n\rangle
\ee
From $V|0\rangle=|0\rangle$ and $|d\rangle\equiv |0\rangle$, we have
\be
e^{ i \xi^{\al} ( 2 \pi)^{\al}   }=1,
\ee
which gives
\be
\xi = (2 \pi)^{ \frac{1}{\al} -1}
\ee
Thus, the operator $V$ is written as
\be
V =  e_{\al} \left[ i (2 \pi)^{ \frac{1}{\al} -1}  \tb \right],
\ee
where we have
\be
V^{\dag} = V^{-1}
\ee
and
\be
V^d = I
\ee
Now let us introduce a unitary translation operator $U$ and its Hermitian adjoint as
\be
U |n \rangle =|n-1\rangle,
 \ee
 and
 \be
U^{\dag} |n\rangle = |n+1\rangle,
\ee
where we assume
 \be
 |n+d\rangle := |n\rangle
 \ee
 Acting $V$ on $U^{\dag} |n\rangle$, we get
 \be
 V( U^{\dag} |n\rangle )
 = V  |n+1\rangle
 = e_{\al} \left[ i (2 \pi)^{ \frac{1}{\al} -1}  \ta_{n+1} \right] |n+1\rangle
 \ee
 Acting  $U^{\dag} |n\rangle$ on $V$, we get
  \be
 U^{\dag}(  V  |n\rangle )
 =  e_{\al} \left[ i (2 \pi)^{ \frac{1}{\al} -1}  \ta_{n} \right]   U^{\dag} |n\rangle
 = e_{\al} \left[ i (2 \pi)^{ \frac{1}{\al} -1}  \ta_{n} \right]  |n+1\rangle
 \ee
 Using Eq.(10), we have
 \be
 V U^{\dag} = q U^{\dag}V,
 \ee
 where
 \be
 q = e_{\al} \left[ i (2 \pi)^{ \frac{1}{\al} -1}  \s \right] = e^{ \frac{ 2 \pi i}{d}}
 \ee
 Similarly, we have
\be
 VU  = q^{-1}  U V
 \ee
 and
 \be
 U^d = (U^{\dag})^d = 1
 \ee
 The commutation relation between the angle operator and unitary translation operator is given by \be
  \tb U \om U \tb = - \s  U
 \ee
 and
 \be
 \tb U^{\dag}\om U^{\dag} \tb = \s  U^{\dag}
 \ee
 Besides, for $r>0$, we have
\be
\tb  (U^{\dag})^ r = (U^{\dag})^r ( X \op r^{1/\al} \s)
\ee
and
\be
\tb U^ r = U^r ( X \om r^{1/\al} \s)
\ee
 Now let us discrete angle  representation of the quantum state $|\psi\rangle$ as
\be
\langle  n | \psi\rangle = \psi (\ta_n)
\ee
{\color{black}By considering the Eq. (22) and the unitary of U,  we obtain}
\be
\langle n | U^{-1} = \langle n-1|
\ee
Multiplying $|\psi\rangle$ from the right we get
\be
\langle n | U^{-1} |\psi\rangle= \langle n-1|\psi\rangle,
\ee
which gives
\be
U^{-1} \psi(\ta_n) = \psi(\ta_{n-1}) =\psi ( \ta_n \om \s)
\ee
Similarly we have
\be
U \psi(\ta_n) = \psi(\ta_{n+1}) =\psi ( \ta_n \op \s)
\ee
In general, for $r>0$,  we have
\be
(U^{-1} )^r \psi(\ta_n) = \psi(\ta_{n-r}) =\psi ( \ta_n \om r^{1/\al} \s)
\ee
\be
U^r  \psi(\ta_n) = \psi(\ta_{n+r}) =\psi ( \ta_n \op r^{1/\al} \s).
\ee

\section{Discrete angular momentum operator and Hermitian Hamiltonian}
Now let us introduce the discrete angular momentum operator. In a discrete theory, we have two types of discrete angular momentum operators,
\be
L_+ = \frac{\h}{i} \lb \frac{ U - I}{\s^{\al}}\rb
\ee
and
\be
L_-= \frac{\h}{i}\lb \frac{ I - U^{-1}}{\s^{\al}}\rb
\ee
{\color{black}Thus, in the discrete angle  representation by substituting Eqs. (38) and (39) in Eq. (42), we have
\begin{eqnarray}
L_+ \psi(\ta_n) &= & \frac{\hbar}{i } \left(\frac{U}{\sigma^{\alpha}} \right) \psi(\theta_{n})-\frac{\hbar}{i } \left(\frac{I}{\sigma^{\alpha}} \right) \psi(\theta_{n})\nonumber\\
&=&\frac{\hbar}{i } \left(\frac{1}{\sigma^{\alpha}} \right) \psi(\theta_{n+1})-\frac{\hbar}{i } \left(\frac{1}{\sigma^{\alpha}} \right) \psi(\theta_{n})\nonumber\\
&=&\frac{\h}{i\s^{\al}} \left(  \psi( \ta_{n+1}) - \psi( \ta_{n})\right) 
\end{eqnarray}
and from Eqs. (38-39) and (43) in the discrete angle  representation, we obtain 
\begin{eqnarray}
 L_- \psi(\ta_n) &=& \frac{\hbar}{i } \left(\frac{I}{\sigma^{\alpha}} \right) \psi(\theta_{n})-\frac{\hbar}{i } \left(\frac{U^{-1}}{\sigma^{\alpha}} \right) \psi(\theta_{n})\nonumber\\
 &=&\frac{\hbar}{i } \left(\frac{1}{\sigma^{\alpha}} \right) \psi(\theta_{n})-\frac{\hbar}{i } \left(\frac{1}{\sigma^{\alpha}} \right) \psi(\theta_{n-1})\nonumber\\
 &=&\frac{\h}{i\s^{\al}} \left(  \psi( \ta_{n}) - \psi( \ta_{n-1})\right) 
\end{eqnarray}
}
Here, we know that neither $L_+$ nor $L_-$ are not Hermitian. Instead, we have
\be
L_+^{\dag} = L_-
\ee
From the non-Hermitian discrete angular momentum operator and its conjugate, we can construct the Hermitian Hamiltonian of the form
\be
H = \frac{1}{2 m R^2 } L_+ L_- + V( \tb),
\ee
where $ m R^2$ denotes {\color{black}an} inertia of moment {\color{black}with} $R$ is a radius of a circle. {\color{black} Also, from Eqs. (42)  and (43), $L_{+}L_{-}$ {\color{black}can be introduced} in the following form
\begin{eqnarray}
L_{+}L_{-}&=&\left[ \frac{\h}{i} \lb \frac{ U - I}{\s^{\al}}\rb \right]\left[\frac{\h}{i}\lb \frac{ I - U^{-1}}{\s^{\al}}\rb \right]  \nonumber\\
&=&- \frac{\h^2}{\s^{2\al}} \left( U + U^{-1} - 2 I\right) 
\end{eqnarray}
 }
Now let us consider a free particle on a circle, {\color{black} whose Hamiltonian by using of Eqs. (47) and (48) is given by}
\be
H = - \frac{\h^2}{ 2 m R^2 \s^{2\al}} ( U + U^{-1} - 2 I)
\ee
{\color{black}As for the time independet Schrodinger equation, the wavefunction of Hamiltonian (49) can be found by applying the translation operators  $U$ and $U^{-1}$ in the discrete angle  representation:}
\be
- \frac{\h^2}{ 2 m R^2 \s^{2\al}} [ \psi( \ta_{n+1}) + \psi (\ta_{n-1}) - 2 \psi ( \ta_n)]
= E \psi(\ta_n)
\ee
{\color{black} Therefore, from Eq. (50) we can get the following form for $(n=2,3,4,...)$:
\begin{eqnarray}
&&\psi(\theta_{2})=\left(2-\frac{2m R^2 \s^{2\al} E}{\h^2} \right) \psi(\theta_{1})-\psi(\theta_{0})\nonumber\\
&&\psi(\theta_{3})=\left[ \left(2-\frac{2m R^2 \s^{2\al} E}{\h^2} \right)^{2}-1\right]  \psi(\theta_{1})-\left(2-\frac{2m R^2 \s^{2\al} E}{\h^2} \right)\psi(\theta_{0})\\
&&\psi(\theta_{4})= \left(2-\frac{2m R^2 \s^{2\al} E}{\h^2} \right)\left[ \left(2-\frac{2m R^2 \s^{2\al} E}{\h^2} \right)^{2}-2\right]  \psi(\theta_{1})-\left[ \left(2-\frac{2m R^2 \s^{2\al} E}{\h^2} \right)^{2}-1\right]\psi(\theta_{0})\nonumber\\
&&\vdots \nonumber
\end{eqnarray}	
where $\psi(\theta_{0})$ and $\psi(\theta_{1})$ are the initial conditions. Then, by choosing of  the following change variable 
\begin{eqnarray}
\cos\xi= \left(1-\frac{m R^2 \s^{2\al} E}{\h^2} \right) \quad,\quad  \sin\xi=\sqrt{1- \left(1-\frac{m R^2 \s^{2\al} E}{\h^2} \right)^{2}}
\end{eqnarray}
and substituting in Eqs. (51), we obtain
}
\be
\psi(\ta_n) = \frac{1}{\sin \xi} \left[ \psi(\ta_1) \sin ( n \xi) - \psi(\ta_0)  \sin (( n -1)\xi) \right],
\ee
or
\be
\psi(\ta_n) = \lb \frac{\psi(\ta_1) - \psi(\ta_0) \cos \xi}{\sin \xi} \rb  \sin ( n \xi) + \psi(\ta_0)  \cos ( n \xi ) ,
\ee
{\color{black}where from Eq. (52), we obtain}
\be
\xi = \tan^{-1} \left[ \frac{ \sqrt{ 1 - \lb 1 - \frac{m R^2 \s^{2\al} E}{\h^2}\rb^2}}{ 1 - \frac{m R^2 \s^{2\al}E}{\h^2}}\right].
\ee
Thus, the energy is given by
\be
E = \frac{(\h d)^2}{ mR^2 (2\pi)^{2\al}}( 1 - \cos \xi),
\ee
which gives the energy bound,
\be
0 \le E \le \frac{2 \h^2}{ m R^2 \s^{2\al}} = \frac{ 2 (\h d)^2}{ mR^2 ( 2 \pi)^{2\al}}
\ee
Thus, the energy is bounded from above in the discrete theory. In the continuum limit, the energy is not bounded from above.
From the cyclic property $\psi(\ta_0)=\psi(\ta_d)$, we have
\be
( \psi(\ta_1) -\psi(\ta_0) \cos \xi) \sin (d \xi) + \psi(\ta_0) \sin \xi \cos ( d \xi) =\psi(\ta_0) \sin \xi
\ee
The quantization rule depend on the value of $d\xi$.
\subsection{ Case of $d \xi = 2 N \pi, ~N\in \mathbb{Z}$}
In this case we have the energy level
\be
E_N = \frac{\h^2 d^2}{ m R^2 (2 \pi)^{2\al}} \lb 1 - \cos \frac{ 2 N \pi}{d}\rb , ~~~N\in \mathbb{Z}
\ee
and corresponding wave function is
\be
\psi_N(\ta_n) = \lb \frac{\psi(\ta_1) - \psi(\ta_0) \cos \lb \frac{2 N \pi}{d}\rb}{\sin \lb \frac{2 N \pi}{d}\rb} \rb  \sin \lb \frac{2 N n  \pi}{d}\rb + \psi(\ta_0)  \cos \lb \frac{2 N n  \pi}{d}\rb .
\ee
\subsection{ Case of $d \xi = 2 N \pi + \frac{\pi}{2}, ~N\in \mathbb{Z}$}

In this case we have
\be
\psi(\ta_1) = \psi(\ta_0) (\cos \xi + \sin \xi)
\ee
Then, we have the energy level
\be
E_N = \frac{\h^2 d^2}{ m R^2 (2 \pi)^{2\al}} \left[ 1 - \cos \lb \frac{  \pi}{d}  \lb 2 N + \frac{1}{2}\rb  \rb\right], ~N\in \mathbb{Z}
\ee
and corresponding wave function is
\be
\psi_N(\ta_n) = \sqrt{2} \psi(\ta_0) \sin \left[ \frac{n \pi}{d} \lb 2 N + \frac{1}{2}\rb + \frac{\pi}{4} \right].
\ee
\subsection{ Case of $d \xi = 2 N \pi + \pi, ~N\in \mathbb{Z}$}

In this case we have $\psi(\ta_0)=0$.
Then, we have the energy level
\be
E_N = \frac{\h^2 d^2}{ m R^2 (2 \pi)^{2\al}} \left[ 1 - \cos \lb \frac{  \pi}{d}  \lb 2 N + 1\rb  \rb\right], ~N\in \mathbb{Z}
\ee
and corresponding wave function is
\be
\psi_N(\ta_n) =  \psi(\ta_0)  \left[ \frac{ \sin \lb \frac{n \pi}{d} \lb 2 N + 1\rb\rb}{
 \sin \frac{\pi}{d} ( 2 N+1)}\right].
\ee

\subsection{ Case of $d \xi = 2 N \pi + \frac{3}{2}\pi, ~N\in \mathbb{Z}$}

In this case we have
\be
\psi(\ta_1) = \psi(\ta_0) (\cos \xi - \sin \xi)
\ee
Then, we have the energy level
\be
E_N = \frac{\h^2 d^2}{ m R^2 (2 \pi)^{2\al}} \left[ 1 - \cos \lb \frac{  \pi}{d}  \lb 2 N + \frac{3}{2} \rb  \rb\right], ~N\in \mathbb{Z}
\ee
and corresponding wave function is
\be
\psi_N(\ta_n) = \sqrt{2} \psi(\ta_0) \cos \left[ \frac{n \pi}{d} \lb 2 N + \frac{3}{2}\rb + \frac{\pi}{4} \right].
\ee

\section{Conclusion}

In this paper we
considered the non-uniformly ( more precisely $\al$-uniformly) distributed grid where we consider the discrete angles are described in terms of the $\al$-addition. Here,
the discrete angles $\ta_n$ were assumed to obey the condition
that the $\al$-difference between two adjacent discrete angles are constant.
With these discrete angles, we defined the angle operator and translation operator. We also discussed the discrete angle representation for the wave function.
Using these, we constructed two types of discrete angular momentum operators and Hermitian Hamiltonian on a circle with $d$ $\al$-distributed discrete angles.
As an example we computed the energy levels
for the free particle on a circle where the wave function is defined in the $d$ $\al$-distributed discrete angles. We found that the energy
is bounded from above in the discrete theory.

\section*{References}
{\color{black}
	
	[1] K. Kowalski, and J. Rembieliński.  Journal of Physics A: Mathematical and General 35, no. 6 (2002): 1405.

	[2] J. Řeháček,  Z. Bouchal, R. Čelechovský, Z. Hradil, and L. L. Sánchez-Soto.  Physical Review A 77, no. 3 (2008): 032110.

	[3] K. Kowalski,  and J. Rembielinski.  arXiv preprint quant-ph/0412150 (2004).

	[4] T. Brzeziński,  J. Rembieliński, and K. A. Smoliński.  Modern Physics Letters A 8, no. 05 (1993): 409-416.

	[5] B. Bahr,  and K.  Liegener.  arXiv preprint arXiv:2101.02676 (2021).

	[6] K .Kowalski, and K. Ławniczak. Journal of Physics A: Mathematical and Theoretical 54, no. 27 (2021): 275302.
	
	[7] B. Bahr, and H. J. Korsch. Journal of Physics A: Mathematical and Theoretical 40, no. 14 (2007): 3959.

	[8] H. A. Kastrup, Physical Review A, 73(5), 052104 (2006).
}


[9] J.Schwinger, Proceedings of the national academy of sciences of the United States
Of America 46, 570 (1960).

[10] A.Vourdas. Reports on Progress in Physics 67, 267 (2004).

[11] T. Santhanam and A. Tekumalla, Foundations of Physics 6, 583 (1976).

{\color{black}
[12] M.Arik and M.Ildes, Progress of Theoretical and Experimental Physics 2016, 041A01
(2016).

[13] D.Bonatsos, C.Daskaloyannis, D.Ellinas  and A.Faessler, Phys.Lett. B 331, 150 (1994).
}

[14]  W. Chung and H.Hassanabadi, Mod. Phys. Lett. B33  (2019) 1950368.

[15] W. Chung and H.Hassanabadi, European Physical Journal Plus 135, Article number: 19 (2020).

\end{document}